%% file: main2.tex
\documentclass{article}
\input{mysnipets}
\usepackage[inline]{enumitem}
\usepackage[version=3]{mhchem}
\usepackage{microtype}
\usepackage{graphicx}
\usepackage{subfigure}
\usepackage{booktabs} 
\usepackage{xspace}
\usepackage{color}
\usepackage{hyperref}
\usepackage{cleveref}
\usepackage{algorithm}
\usepackage[noend]{algpseudocode}
\usepackage[inline]{enumitem}
\usepackage[sortcites,
backend=biber,
hyperref=true,
style=authoryear,
language=auto,
maxbibnames=20,
eprint=false]{biblatex}
\algrenewcommand\algorithmicdo{:}

\title{Graph Energy-based Model for Substructure Preserving Molecular Design}
\author{Ryuichiro Hataya$^{\star,\dagger}$~~ Hideki Nakayama$^{\star}$~~ Kazuki Yoshizoe$^{\dagger}$\\
$^\star$The University of Tokyo ~~~~ $^\dagger$RIKEN AIP}
\date{}

\newcommand{\gem}{\texttt{GEM}\xspace}

\begin{document}
\maketitle

\begin{abstract}
It is common practice for chemists to search chemical databases based on substructures of compounds for finding molecules with desired properties.  The purpose of de novo molecular generation is to generate instead of search.  Existing machine learning based molecular design methods have no or limited ability in generating novel molecules that preserves a target substructure.  Our Graph Energy-based Model, or GEM, can fix substructures and generate the rest.  The experimental results show that the GEMs trained from chemistry datasets successfully generate novel molecules while preserving the target substructures.  This method would provide a new way of incorporating the domain knowledge of chemists in molecular design.
\end{abstract}

\section{Introduction}\label{sec:introduction}

Discovering novel molecules is important but costly and time-consuming. Machine learning based molecular design is expected to remedy this problem by \emph{virtual screening} or \emph{de novo design}: filtering or generating promising candidates from the prohibitively large number of potential compounds. Our focus in this paper is the latter, designing novel molecules.

Most recent de novo design methods use deep generative methods, for instance, GANs and VAEs, to produce graphs \parencite{simonovsky2018,decao2018,jin2018} or string representations (SMILES, \cite{weininger1988smiles}) that is used to describe molecules \parencite{gomez2018automatic,kusner2017grammar}.
These methods help to find candidate compounds with preferable benchmark properties otherwise impossible. 
Despite such advantages, most of the generated molecules are far from what chemists are actually looking for \parencite{gao2020}.
Incorporating substructures with known properties and availability as chemists do may circumvent this problem. 
However, it is either impossible or difficult for existing molecular design methods to incorporate such prior knowledge.

\begin{figure}
    \centering
    \includegraphics[width=\linewidth]{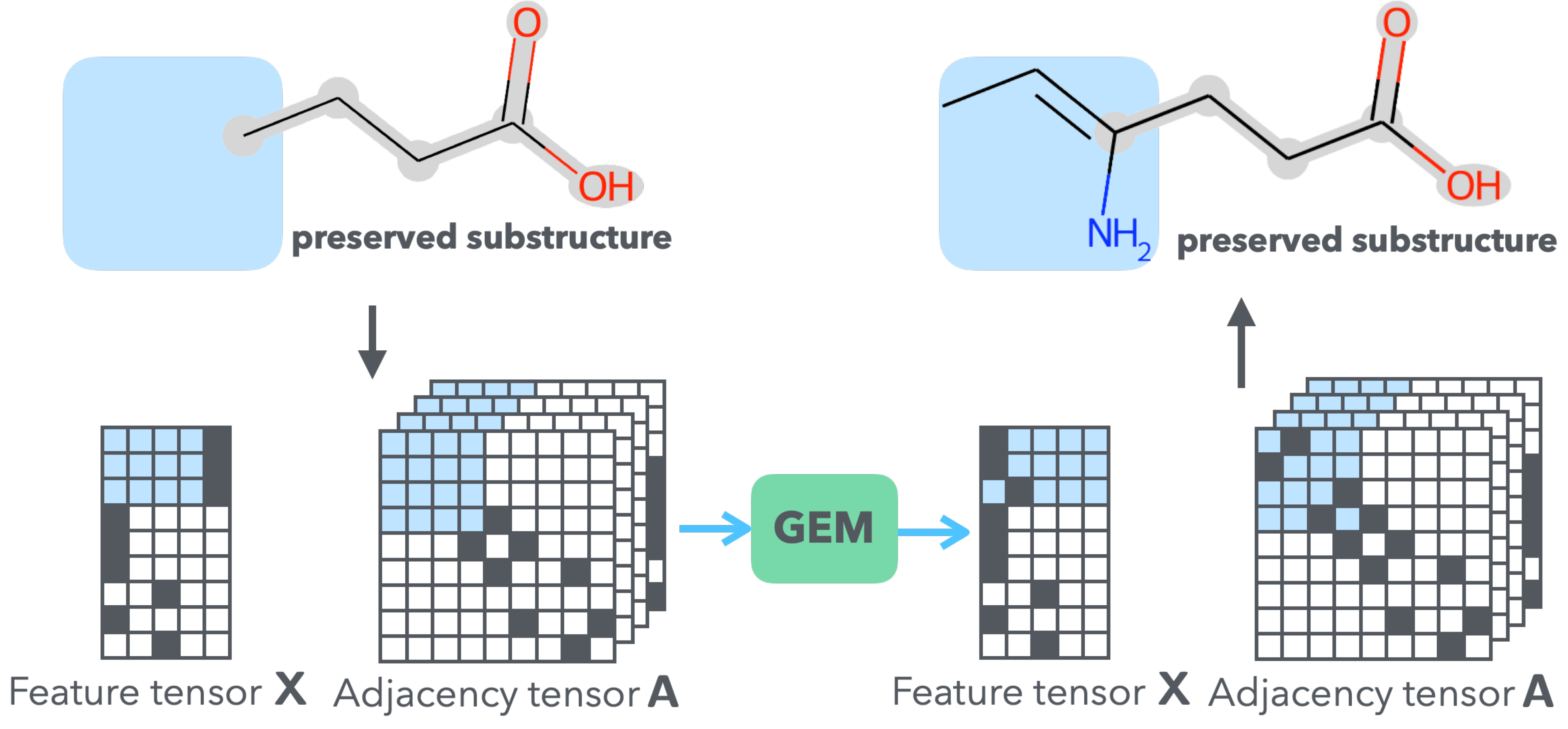}
    \caption{The proposed model, \gem, can generate a molecular graph by preserving its substructure(s), highlighted by gray, and complimenting the rests, highlighted by light blue.}
    \label{fig:overview}
\end{figure}

To this end, we propose \emph{Graph Energy-based Model} (\gem), an energy-based approach for molecular graph generation. \gem can generate molecules including specified substructures (see \Cref{fig:overview}). Energy-based  models (EBMs\footnote{Note that ``energy'' in this context is not ``energy'' in chemistry.} \cite{lecun2006})  are generative models that estimate a density of each data point by Boltzmann-Gibbs distribution with a scalar \emph{energy} function $E$ as $p(\cdot)\propto\exp(-E(\cdot))$. Recently, several studies show that EBMs can generate high-quality images and audio as other deep generative models \parencite{grathwohl2020,song2020,chen2020}.

Inspired by these recent advances of EBMs in image and audio generation tasks, we propose an EBM approach for the molecular graph generation. Unlike the image and audio domains, the graph representation is discrete and constrained, where the techniques in the advances of EBMs are not applicable. To overcome this, \gem uses dequantization and gradient symmetrization. These modifications only slightly change the input graph representations, and off-the-shelf models can be used as generative models.

We empirically demonstrate that \gem can generate molecular graphs while preserving substructure constraints. Additionally, \gem can design molecules with desired properties, such as drug-likeliness, including specified substructures.

\section{Background}\label{sec:background}

\subsection{De Novo Molecular Design}

Novel compound generation using machine-learning requires a way of representing molecules by a data structure.  String-based and graph-based are the two popular approaches for representing molecules.  Many recent studies that apply deep neural networks for molecular design rely on an ASCII string format, Simplified Molecular-Input Line-Entry System (SMILES), which is popularly used in chemistry for describing molecules.  In combination with SMILES based representations, various techniques were used, such as the variations of  VAE~\parencite{gomez2018automatic,kusner2017grammar}, RNN~\parencite{segler2018rnn,yang2017chemts}, and GAN~\parencite{guimaraes2018organ}.

Another popular approach is to use graph-based representation, which seems to be more natural for describing molecules.  Based on the advances in Graph Neural Networks~\parencite{gori2005}, several graph-based approaches, such as GraphVAE~\parencite{simonovsky2018}, JT-VAE~\parencite{jin2018}, and MolGAN~\parencite{decao2018}, outperformed string-based methods in some of the metrics.  Mol-CycleGAN~\parencite{maziarka2020} applied CycleGAN to the latent space of JT-VAE.  Application of graph RNN (MolecularRNN by ~\cite{popova2019}) and the flow model (GraphNVP by ~\cite{kaushalya2019}) also reported their advantages.

Apart from string and graph-based methods, there are other notable studies such as the following.  \cite{kajino2019hypergraph} correctly handles the chemical constraints by hypergraph grammar and assures 100\% validity.  \cite{zhou2019moldqn} defined molecule modification as a Markov decision process and achieved good scores in benchmarks by limiting the type of atoms.  \cite{yoshikawa2018evol} applied grammatical evolution to the problem and succeeded in generating novel molecules.  These approaches have different characteristics than string or graph-based methods and have future potential.

However, none of the previous studies have focused on a substructure preserving generation.  Therefore, the effectiveness of compounds obtained from existing studies is too strongly dependent on the evaluation functions' quality, \eg, penalized logP or QED (see section \ref{sec:experimental_settings}).  In this paper, we propose to use another generative model, an energy-based model, to realize molecular generation while fixing specified substructures.

\subsection{Energy-based Models}

Energy-based models (EBMs, \cite{lecun2006}) estimate probability densities using Boltzmann-Gibbs distributions as

\begin{equation}
    p(\vz)=\frac{\exp(-E(\vz))}{\int_{\vz\in\gZ}\exp(-E(\vz))\mathrm{d}\vz},
\end{equation}

\noindent where $\vz$ is a datum in a certain open set $\gZ$ in a finite-dimensional real space, and $E:\gZ\rightarrow\R$ is a function called energy function. Though the RHS's denominator is usually intractable, EBM methods allow sampling without explicitly obtaining it. High flexibility of the design of $E$ allows wide range of applications, including protein structure prediction \parencite{ingraham2019,du2020} and conformation prediction \parencite{mansimov19}. 

EBMs are also applied to high-resolution image generation, such as \parencite{grathwohl2020,song2020,du2020b}, and high-quality audio wave generation \parencite{chen2020}, which show comparable performance with other popular deep generative models, such as GANs and VAEs. Especially, \cite{grathwohl2020} used a standard image classifier as the energy function, which inspires us to use an energy-based approach for graph generation. However, these image and audio data generations are on concrete domains, different from (molecular) graphs on discrete space.

Exceptionally, \cite{niu2020} applied EBMs to graphs by modeling the score function $\nabla_\vz E$, \ie, using score matching \parencite{hyvarinen05a}, to generate adjacency matrices. In contrast, our approach, \gem, directly uses energy function $E$ and generates molecular graphs with multiple node and edge types.

\section{Graph Energy-based Models}

This section describes our proposed approach, Graph Energy-based Models, or \gem in short.

\begin{figure}
    \centering
    \includegraphics[width=\linewidth]{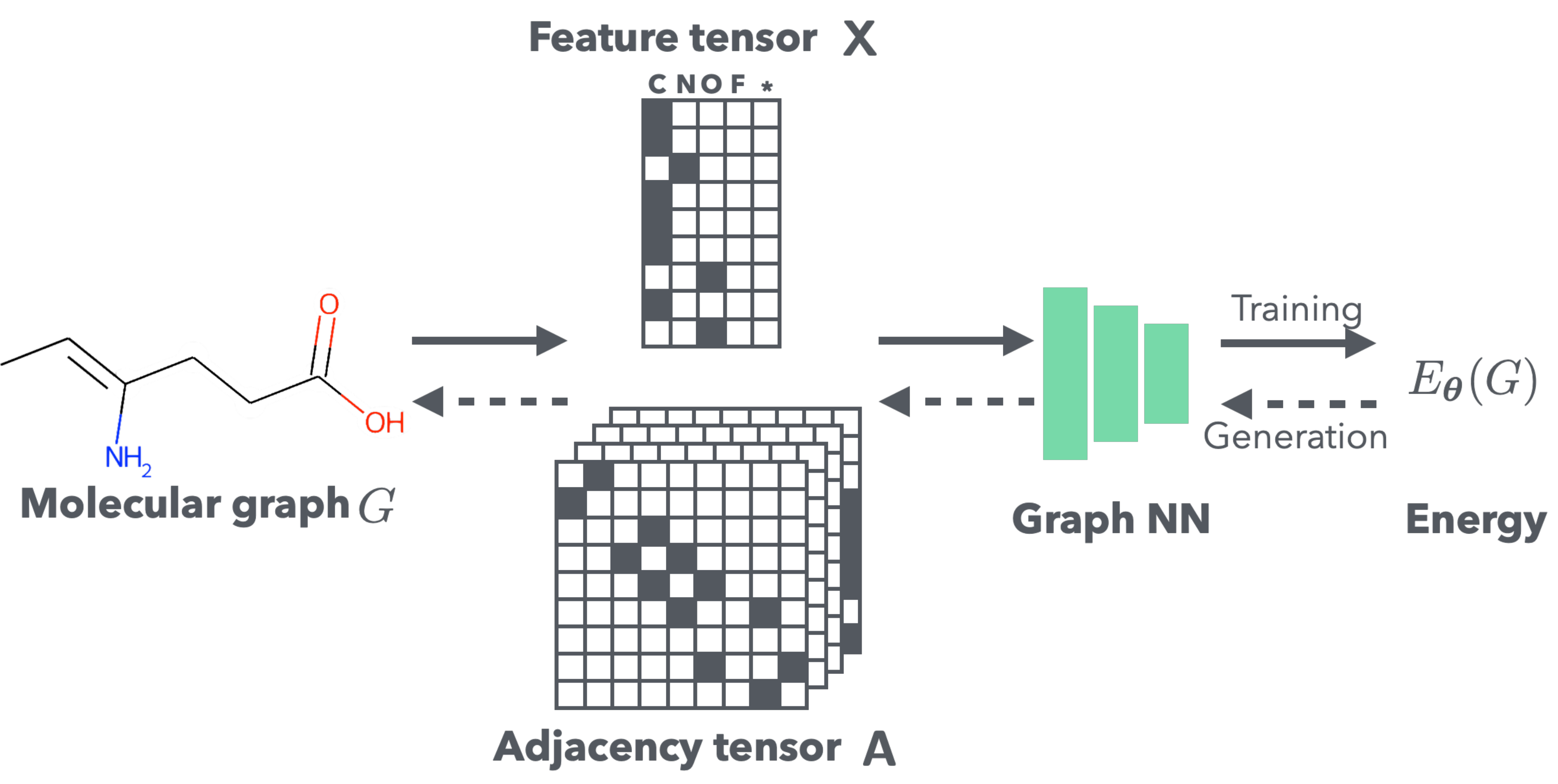}
    \caption{During training, \gem learns to assign lower energy to molecules in a dataset and higher energy to generated ones including invalid graphs. For property-targeted generation, molecules with desired properties are expected to have lower energy. \gem generates molecular graphs with lower energy using MCMC, which are expected to be valid molecules.}
    \label{fig:gem}
\end{figure}

\subsection{Notations}

A molecular graph $G$ is an undirected graph depicted by a pair of tensors: a feature tensor $\tX\in\{0, 1\}^{N\times \#\gM}$ and an adjacency tensor $\tA\in\{0, 1\}^{N\times N\times \#\gB}$. The feature tensor $\tX$ represents atoms in the molecule, and the adjacency tensor represents bonds among them. $N$ is the maximum number of atoms in molecules in a dataset, $\gM$ is a set of considered atoms, \eg, $\gM=\{\ce{C}, \ce{N}, \ce{O}, \ce{F}, \text{virtual node}\}$, and $\gB$ is the set of bond types, \ie, $\gB=\{\text{single}, \text{double}, \text{triple}, \text{virtual bond}\}$. ``virtual node'' and ``virtual bond'' are used for padding in case the number of atoms in a given molecule is smaller than $N$.

For each triplet of $(N, \gM, \gB)$, there is a set of valid molecular graphs $\gG=\gG_{(N, \gM, \gB)}$. Validity includes the symmetry of adjacency tensor slices: $\tA_{:,:,b}$ is a symmetric matrix for $b=1,2,\dots,\#\gB$. Practically, we use datasets $\gD\subset\gG$.

\subsection{Generating Graphs by EBMs}\label{sub:gemeration_by_ebm}

We propose to generate a novel molecule by using an energy function $E_\vtheta: \gG\to\R$, parameterized by a real vector $\vtheta$. Specifically, we use a graph neural network to represent this parameterized function. The energy function is expected to assign smaller values to valid molecules and higher values to invalid ones. This energy function determines a Boltzmann-Gibbs distribution

\begin{equation}\label{eq:ebm}
    p_\vtheta(G) = \frac{\exp(-E_\vtheta(G))}{\sum_{G'\sim\gG}\exp(-E_\vtheta(G))},
\end{equation}

\noindent from which molecules are expected to be sampled with a high probability. If graphs are continuous, we can sample graphs from this distribution by using stochastic gradient Langevin dynamics (SGLD, \cite{welling11}):

\begin{align}
    \tX^{(t+1)} &= \tX^{(t)} + \frac{\alpha_t}{2} g_\tX(\tX^{(t)}, \tA^{(t)}) + \sqrt{\alpha_t}\vepsilon_{\tX}, \label{eq:gen_x} \\
    \tA^{(t+1)} &= \tA^{(t)} + \frac{\alpha_t}{2} g_\tA(\tX^{(t)}, \tA^{(t)}) + \sqrt{\alpha_t}\vepsilon_{\tA}, \label{eq:gen_adj}
\end{align}

\noindent where $\alpha_t\in\R^+$ is a step size, $g_\tX=\nabla_\tX E_\vtheta$ and $g_\tA=\nabla_\tA E_\vtheta$ are score functions, and $\vepsilon_{\tX}$ and $\vepsilon_{\tA}$ are standard normals. This generation (\Cref{eq:gen_x,eq:gen_adj}) can also be achieved by directly estimating $g_\tX$ and $g_\tA$ as \parencite{niu2020}. $\tX^{(0)}$ and $\tA^{(0)}$ are sampled from a uniform distribution on $[0, 1]$. The distribution of $G^{(\infty)}=(\tX^{(\infty)}, \tA^{(\infty)})$ is asymptotically equal to $p_\vtheta(G)$, and we assume that this property can be approximated with finite steps with a small constant state size, \ie, $\alpha_t=\alpha$, following the literature.

Actually, simply applying \Cref{eq:gen_x,eq:gen_adj} does not work in our case, because they do not consider the following requirements:
\begin{enumerate*}
    \item $\tX$ and $\tA$ are discrete, and
    \item slices of $\tA$ is symmetric.
\end{enumerate*}
To fix the first issue, we relax the domains of $\tX$ and $\tA$ to be $(0, 1)^{N\times \#\gM}$ and $(0, 1)^{N\times N\times \#\gB}$. For discrete tensors from datasets, we modify them by using \emph{dequantization} and applying softmax function along the last axes. Dequantization is a technique used in \cite{kaushalya2019}, which adds random values to the tensor elements:

\begin{equation}
    \tX\leftarrow \tX+c\tU_{\tX},~~ \tA\leftarrow \tA+c\tU_{\tA},
\end{equation}

\noindent where $c\in(0, 1)$ is a scaling parameter, and $\tU_{\tX}, \tU_{\tA}$ are uniform noise on $(0, 1)$. We set $c=0.9$ in the experiments.

To avoid sampled adjacency tensors being asymmetric, we sample $\tA^{(0)}$ and $\vepsilon_\tA$ from symmetric distributions, where

\begin{equation*}
    (\tA^{(0)})_{i,j,b}=(\tA^{(0)})_{j,i,b},~~ (\vepsilon_{\tA})_{i,j,b}=(\vepsilon_{\tA})_{j,i,b},
\end{equation*}

\noindent for $i,j\in\{1,2,\dots,N\}$ and $b\in\{1,2\dots,\#\gB\}$. Additionally, the score function $g_\tA$ needs to be symmetric, which we will describe in the next section.

\subsection{Symmetrize Gradient of Adjacency Tensor}\label{sub:model}

We use a neural network based on Relational GCN (RGCN) \parencite{Schlichtkrull2018} as an energy function. RGCN is a graph convolutional neural network for graphs with multiple edge types. For each graph $G=(\tX, \tA)$, the $l$th RGCN layer processes node representation $\tH_l\in\R^{N\times C}$ as

\begin{equation}\label{eq:rgcn}
    \tH_{l+1} = \sigma\left( \tH_{l} W_{l}^{(0)}+\sum_{b=1}^{\#\gB}\tA_{:,:,b}\tH_{l}W_{l}^{(b)} \right),
\end{equation}

\noindent where $\tH_0=\tX$, $W_{l}^{(0)}, W_{l}^{(b)}\in \R^{C\times D}$ are learnable parameters, $\sigma$ is a nonlinear activation function, and $C, D$ are input and output feature dimensions. After several RGCN layers, a graph-level representation is obtained by the aggregation of \parencite{li2016}. This representation is transformed into a scalar value $E_\vtheta(G)$ by a multi layer perceptron.

Crucially, with this energy function, the score function $g_\tA$ is asymmetric. Indeed, by focusing on the first layer of RGCN layers and ignoring the nonlinear activation for simplicity, we obtain a Jacobian tensor of

\begin{equation}\label{eq:rgcn_grad}
    \frac{\partial \tH_1}{\partial (\tA)_{i,j,b}}
  = \frac{\partial \tA_{:,:,b}\tX W_b^{(1)}}{\partial (\tA)_{i,j,b}}
  = \mJ^{(i,j)}\tX W_b^{(1)},
\end{equation}

\noindent where $\mJ^{(i,j)}$ denotes a single entry matrix of $1$ at $(i,j)$ and $0$ elsewhere \parencite{matrixcookbook}. This gradient is not symmetric for each $b$. To remedy this, we modify \Cref{eq:rgcn} as 

\begin{equation}\label{eq:rgcn_mod}
    \tH_{l+1} = \sigma\left( \tH_{l} W_{l}^{(0)}+\sum_{b=1}^{\#\gB}\frac12(\tA_{:,:,b}+\tA_{:,:,b}^{\top})\tH_{l}W_{l}^{(b)} \right).
\end{equation}

Though this modification does not change the output because each $\tA_{:,:,b}$ is symmetric by definition, now the Jacobian tensor is also symmetrized as

\begin{equation}
    \frac{\partial \tH_1}{\partial (\tA)_{i,j,b}}=\frac12(J^{(i,j)}+J^{(j,i)})\tX W_b^{(1)},
\end{equation}

\noindent from which we can deduce $\displaystyle \frac{\partial E_\vtheta}{\partial (\tA)_{i,j,b}}=\frac{\partial E_\vtheta}{\partial (\tA)_{j,i,b}}$, the symmetry of the score function $g_\tA$.
Practically, the modification of \Cref{eq:rgcn_mod} can be separately done before the forward pass of the model, which means the actual modification to the off-the-shelf models is minimum. In the experiments, we use the abovementioned RGCN variant, which is also used in other graph-based molecular generation methods \parencite{decao2018,kaushalya2019}.

\subsection{Training of \gem}\label{sub:training}

\gem is trained to maximize likelihood of $p_\vtheta$ defined in \Cref{eq:ebm}. Equivalently, this objective minimizes the Kullback-Leibler divergence between data and model distribution $\KL{p_\gD}{p_\vtheta}$. To optimize the energy function $E_\vtheta$, we can use stochastic gradient of

\begin{equation}\label{eq:training}
    \nabla_\vtheta\E_{\gD}\left[\log p_\vtheta(G)\right] 
    = \E_{\gD}\left[\nabla_\vtheta E_\vtheta(G)\right] - \E_{p_\vtheta(G')}\left[\nabla_\vtheta E_\vtheta(G')\right].
\end{equation}

At the LHS's second term, samples from the model $G'\sim p_\vtheta(G')$ are used. As discussed in \Cref{sub:gemeration_by_ebm}, we use a finite step of SGLD to approximate this sampling, resulting in diverged samples from the model distribution. To remedy this problem, we use the persistent contrastive divergence (PCD, \cite{tielemen08}), which reuses the past generated samples.

We summarize the training procedure in \Cref{alg:gem}. Additionally, we found that penalizing $\{E_\vtheta(G)\}^2$ improves empirical performance.

\begin{algorithm}
    \caption{Training of \gem}
    \label{alg:gem}
    
    \begin{algorithmic}
        \Statex Energy function: $E_\vtheta$
        \Statex Reinitialization probability: $\rho$
        \While{not converge}
            \State Sample $G=(\tX_G, \tA_G)$ from dataset $\gD$
            \State Sample $(\tX_B, \tA_B)$ from PCD buffer $B$ with probability $1-\rho$, otherwise sample $(\tX_B, \tA_B)$ from uniform distribution on $(0, 1)$
            \State SGLD steps in \Cref{eq:gen_x,eq:gen_adj}:\hfill$(\tX_B, \tA_B)\mapsto(\tX'_B, \tA'_B)$
            \State Compute energy loss\hfill $\ell_\vtheta=\nabla_\vtheta E_\vtheta(\tX_G, \tA_G)-\nabla_\vtheta E_\vtheta(\tX'_B, \tA'_B)$
            \State \textcolor[gray]{0.3}{For property-targeted generation, add regression loss \\\hfill$\ell_\vtheta\leftarrow \ell_\vtheta+\abs{E_{\vtheta}(G)-y_G}^2$, where $y_G$ is a property value of $G$}
            \State Add $(\tX'_B, \tA'_B)$ to $B$
            \State Update $\vtheta$ with stochastic gradient $\nabla_\vtheta \ell_\vtheta$
        \EndWhile
    \end{algorithmic}
\end{algorithm}

\subsection{Generation by \gem}

Once the energy function $E_\vtheta$ is trained, \gem can generate molecular graphs using SGLD in \Cref{eq:gen_x,eq:gen_adj} unconditionally. Additionally, \gem can design molecules while preserving substructures and optimizing desired properties.

\subsubsection*{Substructure Preserving Generation}

\begin{figure}
    \centering
    \includegraphics[width=0.6\linewidth]{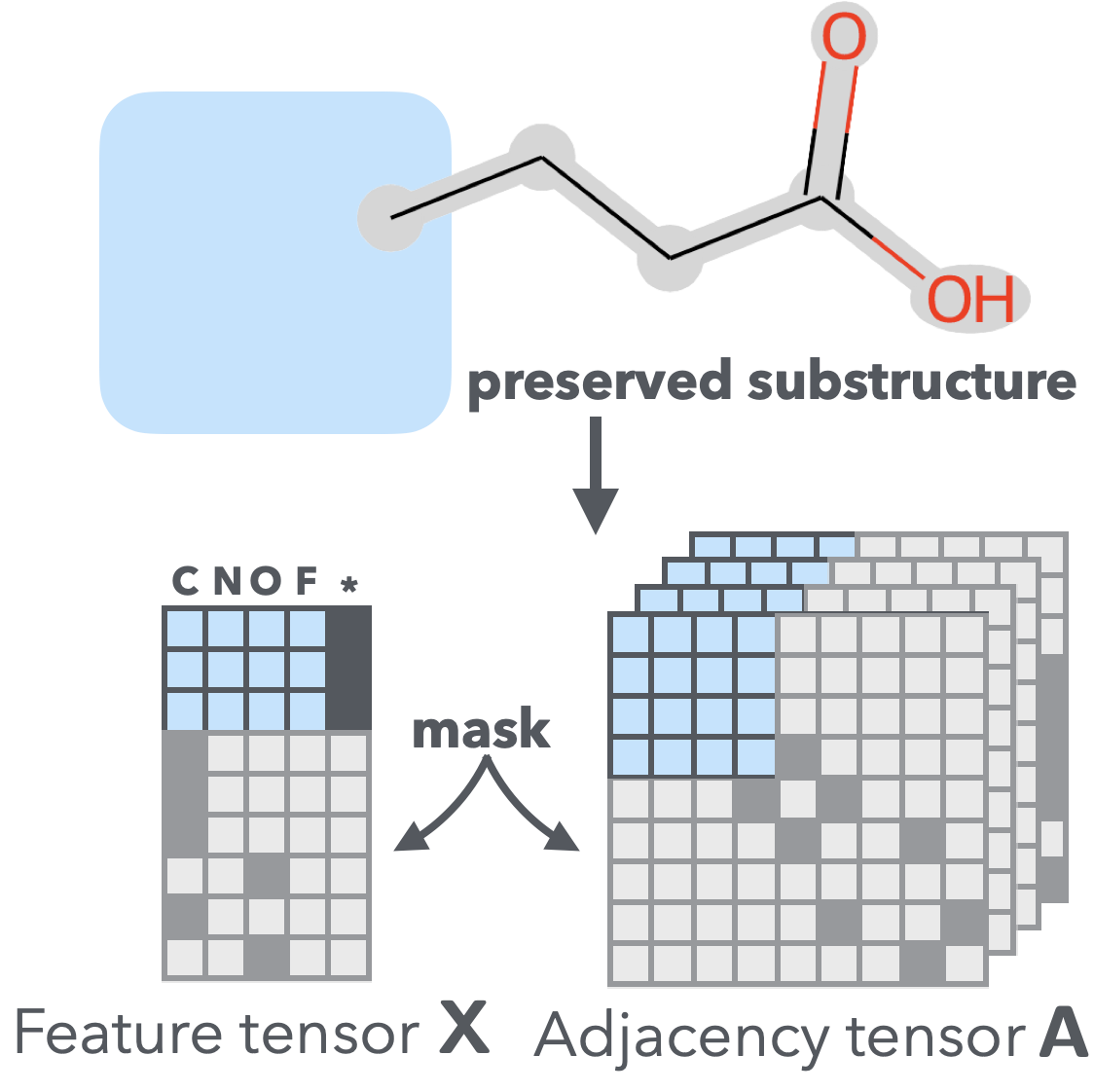}
    \caption{\gem enables substructure preserving generation by masking feature and adjacency tensors (masked by light gray) and updating the rests parts (highlighted by light blue).}
    \label{fig:mask}
\end{figure}

The most characteristic ability of \gem is to generate molecular graphs while fixing specified substructures.
Because \gem samples molecular graphs in the input space by SGLD, this ability is achieved by updating parts of graph representation (see also \Cref{fig:overview,fig:mask}).

To fix substructures, we apply masks to both feature and adjacency tensors. Suppose the number of atoms in a given substructure is $S<N$, where $N$ is the maximum number of atoms in molecules of a dataset. Because \gem is permutation invariant to an input representation, we can re-index atoms in the substructure to $1, 2, \dots, S$ such that the $S$th atom to be connected with the rest part, without loss of generality. Then, we use a mask to update only a part of the feature tensor corresponding to $S+1, S+2, \dots, N$th atoms and fix the atoms in the substructure. Similarly, we only update connections among $S, S+1, \dots, N$th atoms and fix the connections among the rests by masking the adjacency tensor. This masking can be extended to appending the rest parts to multiple atoms.

\subsubsection*{Property-targeted Generation}

\gem explicitly estimates energy function $E_\vtheta$. One of the most appealing benefits of this modeling is property-targeted molecular generation, by regarding an objective property as energy. Specifically, lower energy is assigned to molecules with desired properties, and \emph{vice versa}. Because \gem samples molecular graphs with lower energy, this assignment brings \gem to generate compounds with desired properties. For this purpose, \gem is trained jointly with a regression loss $\E_{G\sim\gD}\abs{E_{\vtheta}(G)-y_G}^2$, where $y_G$ is a property, such as drug-likeliness, of a molecule $G$.

\section{Experiments}

\subsection{Experimental Settings} \label{sec:experimental_settings}

\subsubsection*{Datasets}
\newcommand{\qm}{\texttt{QM9}\xspace}
\newcommand{\zinc}{\texttt{ZINC-250k}\xspace}

We used \qm \parencite{qm9} and \zinc \parencite{zinc} as datasets $\gD$. \qm and \zinc contain $1.3\times10^5$ and $2.5\times10^5$ molecules, respectively. Following the preprocessing protocols in \cite{kaushalya2019}, we kekulize each molecule in each dataset and ignore hydrogens as the SMILES format. As a result, the maximum number of atoms in a molecule $N$ is $9$ for \qm9 and $38$ for \zinc. The number of atom types $\#\gM$ including the virtual node is $5$ for \qm and $10$ for \zinc. The number of bond types $\#\gB$ is 4, namely $\gB=\{\text{single}, \text{double}, \text{triple}, \text{virtual bond}\}$, for both datasets. We also followed the data split of \cite{kaushalya2019}.

\subsubsection*{Implementation Details}

We used PyTorch v1.7 \parencite{pytorch} for model implementation, chainer-chemistry v0.7 \footnote{\url{https://github.com/chainer/chainer-chemistry}} for data preprocessing, and RDKit  v2020.09  \footnote{\url{https://www.rdkit.org}} for handling molecule information.

Each input feature tensor is embedded in 16-dimensional space and processed by a two-layer RGCN of 128 hidden dimensions. Its output is aggregated in a 256-dimensional space and converted to scalar energy by an MLP of $(1024, 512)$ hidden units. The hyperbolic tangent function is used as an activation function, and the sigmoid function $\varsigma(x)=\{1+\exp(-x)\}^{-1}$ is applied to the final output that restricts the range to $[0, 1]$.

We trained \gem using Adam \parencite{kingma2015} with a learning rate of $1.0\times10^{-4}$ for $30$ epochs. 
For SGLD, we set a step size $\alpha$ to $1.0\times10^{-4}$ and the number of steps to $40$. Following \cite{grathwohl2020,du2020b}, we set the buffer size of PCD to $10^4$ and the reinitialization probability $\rho$ (see \Cref{alg:gem}) to $5.0\times10^{-2}$, and reduced the effect of additive noise by multiplying $0.1$ to the standard deviation as common practice. For SGLD, we used an exponential moving average of the model with a decay rate of $1.0\times10^{-3}$ for the stability.

To generate molecular graphs, we used SGLD of step size of $1.0\times10^{-1}$ for \qm and $1.0\times10^{-2}$ for \zinc, and the number of steps of $10^3$. Adding noise in \Cref{eq:gen_x,eq:gen_adj} sometimes turns once generated valid molecular graphs into invalid ones. Therefore, we record all valid graphs generated at each step. We discarded the graphs generated during the first 100 steps to reduce the effects of initial states.

\subsubsection*{Objective Properties}

For property-targeted generation, we use the following commonly used properties:

\begin{description}[nosep]
    \item [Penalized logP (solubility):] hydrophobicity, namely the logarithm of octanol-water partition coefficient penalized by synthetic accessibility and ring penalty, and
    \item [Drug-likeliness (QED):] measure of drug likeliness, specifically the quantitative estimate of drug-likeness \parencite{qed}.
\end{description}

We used RDKit to compute these properties of each compound. In the experiments, we normalize these measures for each molecule into $[0, 1]$ as $0$ to be a favorable property value because \gem generates molecular graphs with lower energy.

\subsection{Substructure Preserving Generation}

\begin{figure}
    \centering
    \includegraphics[width=0.7\linewidth]{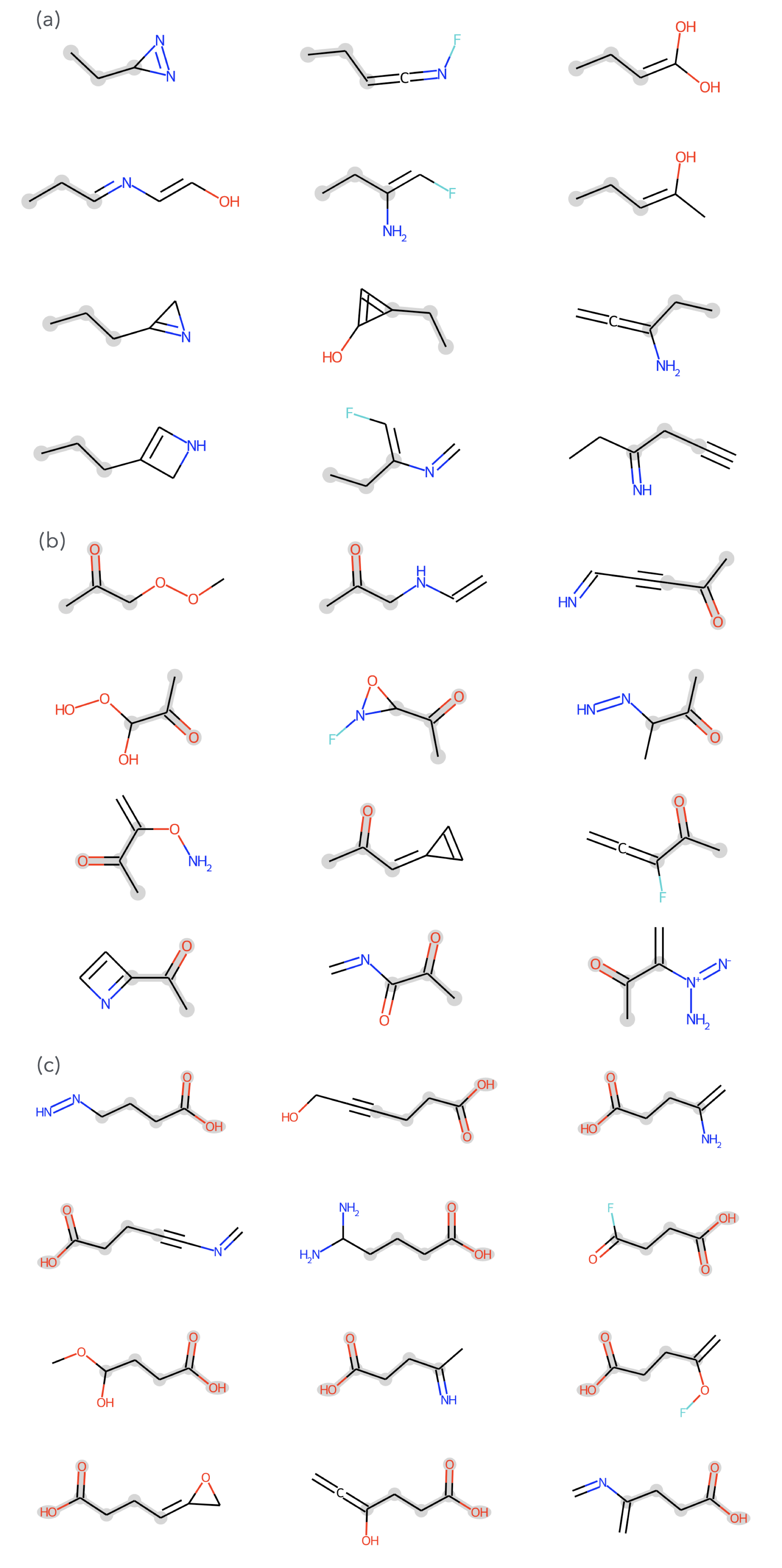}
    \caption{Randomly selected molecules of substructure preserving generation. Conditioned substructures are (a) propane \texttt{CCC}, (b) acetone \texttt{CC(=O)C}, and (c) butanoic acid \texttt{CCCC(=O)O}, which are highlighted by light gray.}
    \label{fig:qm}
\end{figure}

\Cref{fig:qm} shows examples of substructure preserving generation using \gem, which is impossible for existing graph-based generative methods. %
As conditioning substructures, we used propane \texttt{CCC}, acetone \texttt{CC(=O)C}, and butanoic acid \texttt{CCCC(=O)O}. Carbon atoms at the edge of each molecule is specified to append generated parts. \gem can successfully generate molecules while preserving specified substructures.
% ...

\begin{figure*}
    \centering
    \includegraphics[width=0.8\linewidth]{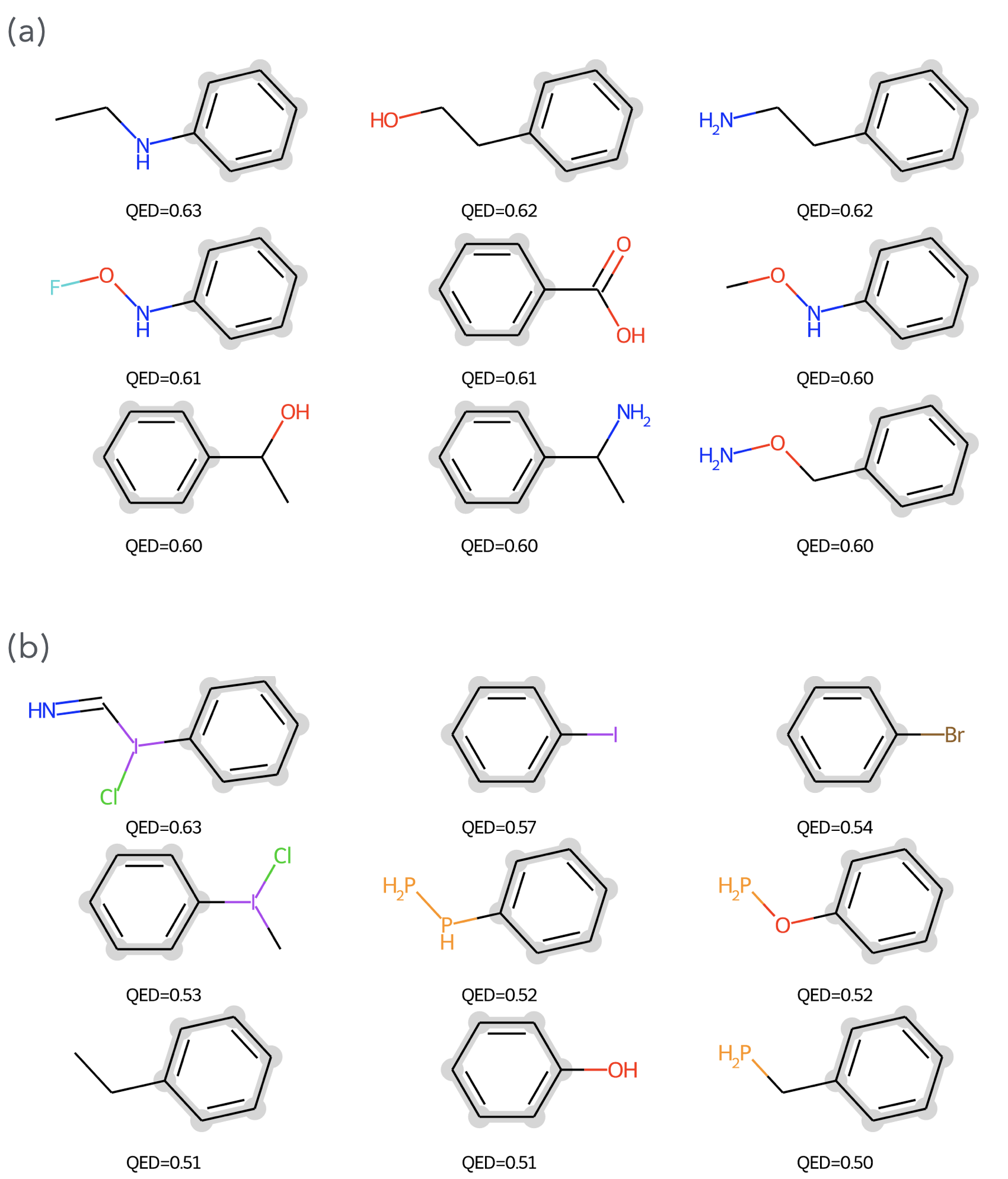}
    \caption{\textbf{Generated molecules while preserving a substructure, benzene \texttt{c1ccccc1}, and optimizing drug-likeliness (QED)}. \gem is trained on \qm for (a) and \zinc for (b). The preserved substructure is highlighted by light gray. Generated molecules with the best QED score are presented. The QED value of the original substructure is $0.44$.}
    \label{fig:qm_conditioned}
\end{figure*}

Substructure preserving generation can be combined with property-targeted optimization. \Cref{fig:qm_conditioned} presents generated molecules by \gem trained on \qm and \zinc with conditioning substructure of benzene \texttt{c1ccccc1}. \gem is trained on each dataset with regression loss to drug-likeliness as presented in \Cref{alg:gem}. As can be seen, \gem generates molecules with improved QED values (up to $0.19$), while preserving the benzene substructure.

\begin{figure}
    \centering
    \includegraphics[width=\linewidth]{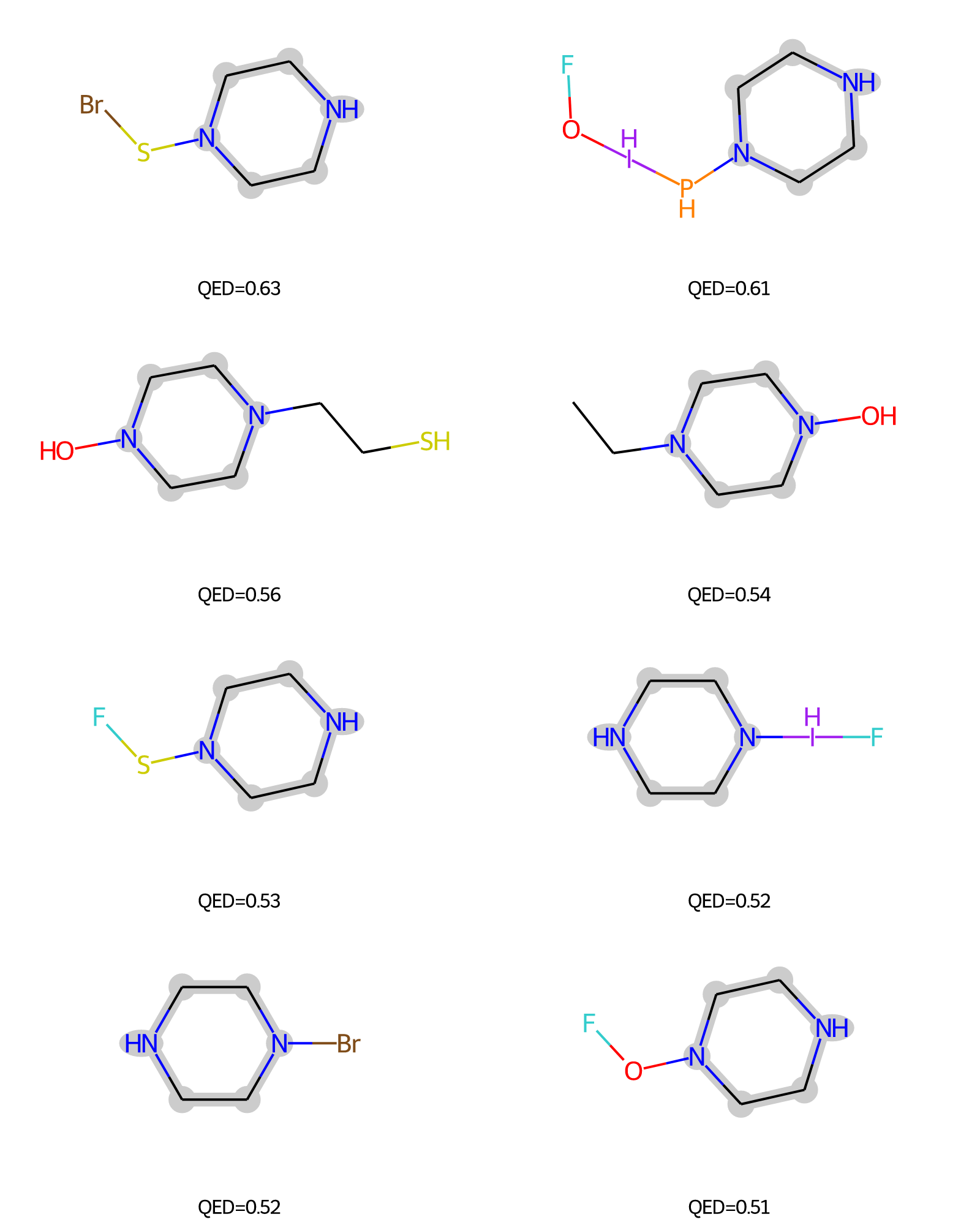}
    \caption{\textbf{Generated molecules while preserving a substructure, piperazine \texttt{C1CNCCN1}, and optimizing drug likeliness (QED).} \gem is trained on \zinc. The preserved substructure is highlighted by light gray. Generated molecules with the best QED scores are presented. QED of the original substructure is $0.40$.}
    \label{fig:zinc_conditioned1}
\end{figure}

\begin{figure}
    \centering
    \includegraphics[width=\linewidth]{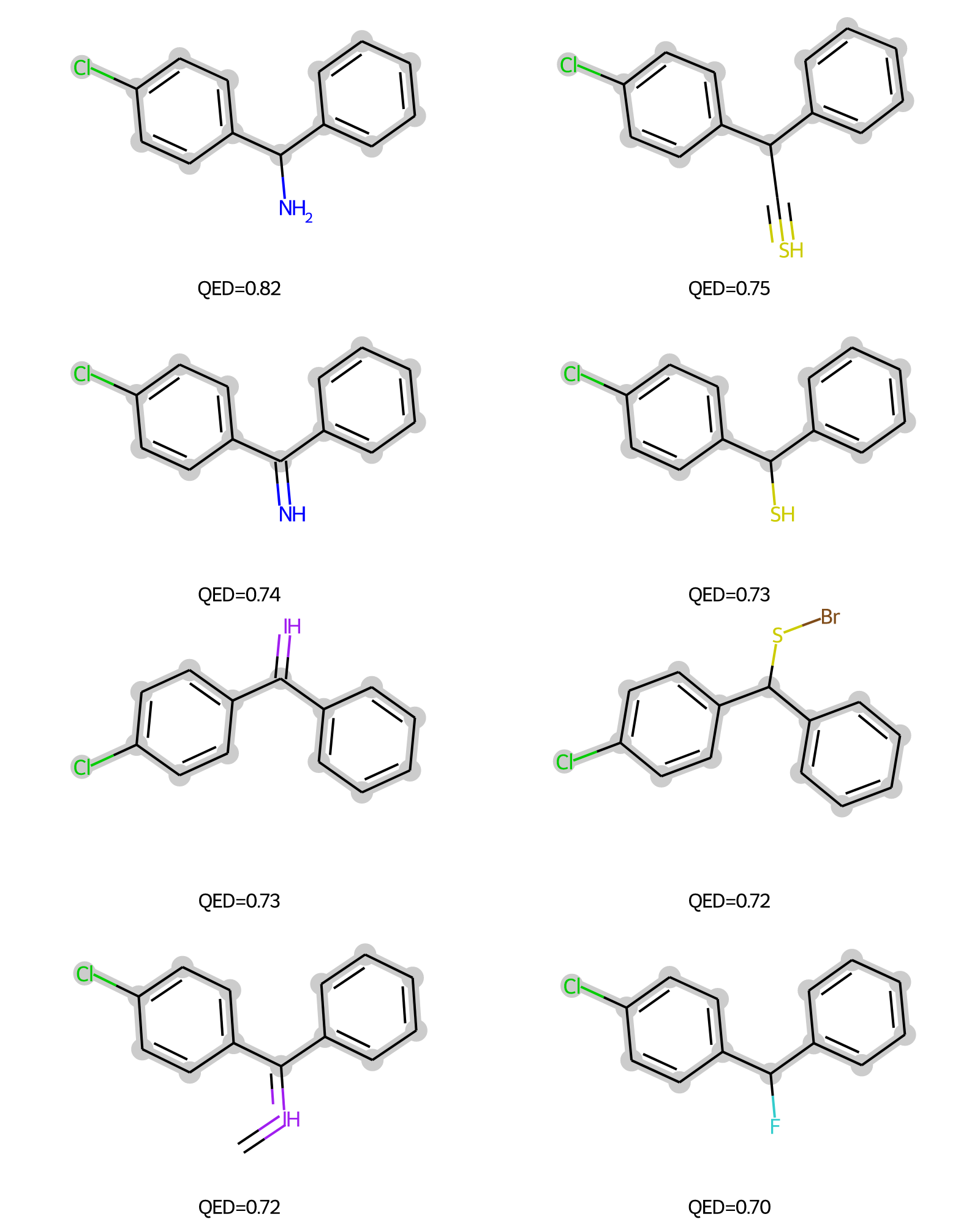}
    \caption{\textbf{Generated molecules while preserving a substructure, \texttt{c1ccc(cc1)Cc2ccc(cc2)Cl}, and optimizing drug likeliness (QED).} \gem is trained on \zinc. The preserved substructure is highlighted by light gray. Generated molecules with the best QED scores are presented. QED of the original substructure is $0.69$.}
    \label{fig:zinc_conditioned2}
\end{figure}

As more complex examples, \Cref{fig:zinc_conditioned1,fig:zinc_conditioned2} show generated molecular graphs conditioned by piperazine \texttt{C1CNCCN1} and 4-Chlorodiphenylmethane \texttt{c1ccc(cc1)Cc2ccc(cc2)Cl}\footnote{Specifically, the molecule used in \cite{compret} is 4-Chlorobenzophenone \texttt{c1ccc(cc1)C(=O)c2ccc(cc2)Cl}. The oxygen atom is removed during synthesis, and thus \texttt{c1ccc(cc1)Cc2ccc(cc2)Cl} is used in our experiments.}, which are used in \parencite{compret} as starting materials of retro-synthesis. \gem is trained on \zinc jointly with regression loss to drug-likeliness. 
For piperazine \texttt{C1CNCCN1} (\Cref{fig:zinc_conditioned1}), we specified both of two nitrogen atoms to append generated parts. Such generation with complex conditioning is almost impossible for SMILE-based methods that may only append generated parts subsequent to given substructures. Contrarily, \gem enjoys high flexibility of substructure-preserving generation.

\section{Discussion}

\subsection{Improving Given Molecules}

\begin{figure}[t]
    \centering
    \includegraphics[width=0.5\linewidth]{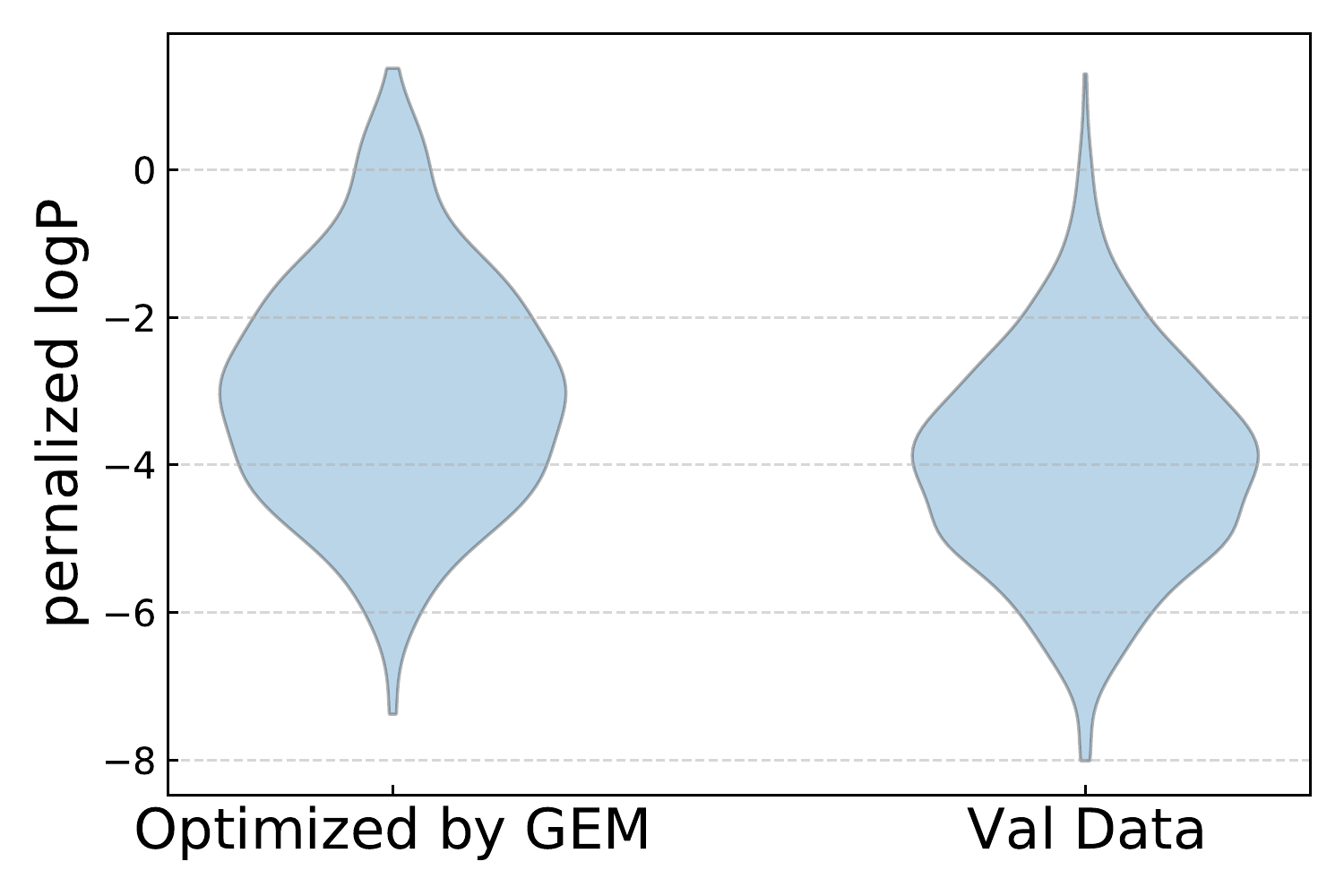}
    \caption{Penalized logP comparison of molecular graphs optimized by \gem from validation data and the original molecules.}
    \label{fig:optimize_data}
\end{figure}

So far, \gem creates compounds from uniform noise of the initial states of generation (\Cref{eq:gen_x,eq:gen_adj}) with fixed substructures. Additionally, \gem can improve given molecules with respect to targeted properties by substituting them for random initial states and removing masks for preservation.
We sampled $10^3$ molecules from the validation set of \qm and optimized penalized logP using \gem. \Cref{fig:optimize_data} compares penalized logP of molecules generated by \gem trained on \qm with regression loss to penalized logP and the original molecules from the dataset. \gem can effectively optimize penalized logP from the original data. Notice that this property improvement can further be integrated into substructure preserving generation.

\subsection{Comparison with Random Generation}

\begin{figure}[t]
    \centering
    \includegraphics[width=0.5\linewidth]{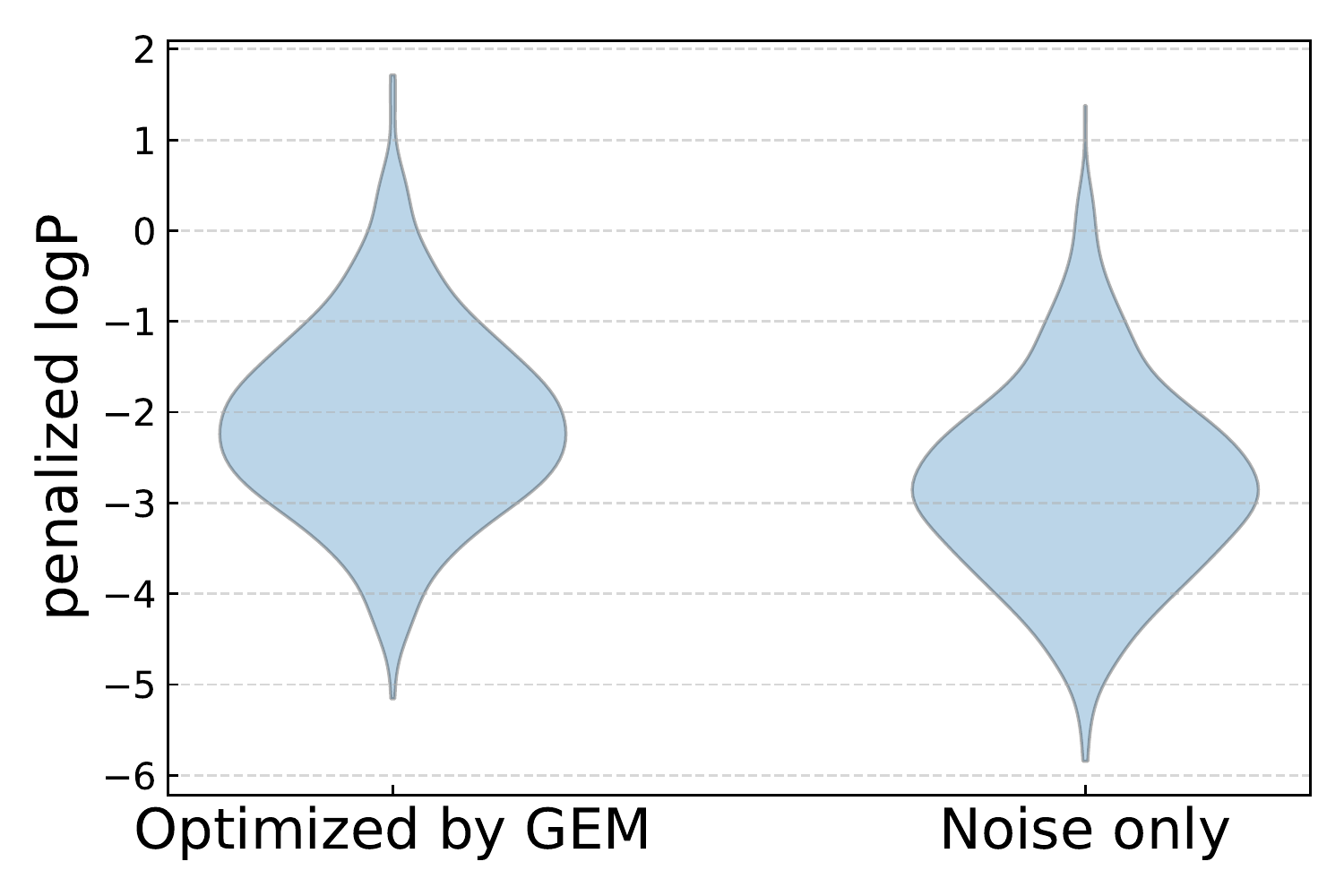}
    \caption{Penalized logP comparison of molecular graphs generated by \gem and noise using a substructure of acetone.}
    \label{fig:compare_with_random}
\end{figure}

We found SGLD (\Cref{eq:gen_x,eq:gen_adj}) ignoring score functions and only adding noise, namely, $\tX^{(t+1)} = \tX^{(t)} + \sqrt{\alpha}\vepsilon_{\tX}, \tA^{(t+1)} = \tA^{(t)} + \sqrt{\alpha}\vepsilon_{\tA}$, can sometimes produce valid molecular graphs. In \Cref{fig:compare_with_random}, we compare penalized logP of molecular graphs generated by \gem and noise using the substructure condition of acetone. In this setting, \gem is trained on \qm jointly to maximize penalized logP. As can be observed, \gem can generate molecular graphs with desired property, penalized logP. These results indicate that generated graphs are sampled from the graph distribution induced from the trained energy function. Additionally, SGLD without noise, \ie, $\tX^{(t+1)} = \tX^{(t)} + \frac{\alpha}{2}g_\tX(\tX^{(t)}, \tA^{(t)}), \tA^{(t+1)} = \tA^{(t)} + \frac{\alpha}{2}g_\tA(\tX^{(t)}, \tA^{(t)})$ failed to produce graphs, which shows the importance of both score functions and noise in \gem.

\subsection{Comparison with Non Substructure-preserving Methods}

We observed that \gem produces less valid graphs when substructures are not specified. As a reference, we compare \gem with other graph-based molecular generation methods. \Cref{tab:comparison} presents scores of validity and novelty of \gem and other approaches, namely GraphNVP \parencite{kaushalya2019},  MolGAN \parencite{decao2018}, and RVAE \parencite{ma2018}, which use the normalizing flow, GAN, and VAE as backend generative methods, respectively. These models generate adjacency tensors in a one-shot manner. For comparison, we generate molecular graphs from 1,000 different random pairs of feature and adjacency tensors. Unlike baseline methods, the metrics of \gem are computed using unique molecules, which disallows duplication. This design may underestimate the ability of \gem. 

\begin{table}
	\centering
	\caption{The results of non substructure-preserving molecular graph generation. All models are trained on \qm. Baseline results are borrowed from the original papers. For \gem, average and standard deviation of five runs are reported.}
	\label{tab:comparison}
	\begin{tabular}{lcc}
	\toprule
	Method                    &  Validity $(\uparrow)$   &  Novelty $(\uparrow)$  \\
	\midrule
	\gem                      &  \hphantom{0}$7.1\pm0.6$   &   $92.4\pm2.9$     \\
	\midrule
	\gem (\texttt{CC(=O)C})   &  $49.2\pm2.4$              &   $89.0\pm0.7$      \\
	\gem (\texttt{CCCC(=O)O}) &  $22.9\pm0.9$              &   $100\pm0.0$      \\
	\midrule
	GraphNVP    &  $83.1\pm0.5$              &  $58.2\pm1.9$   \\
	MolGAN      &  98.1                      &  94.2                \\
	RVAE        &  96.6                      &  97.5                \\
	\bottomrule
	\end{tabular}
\end{table}

\subsection{Future Direction}

\Cref{tab:comparison} shows limited validity of generated molecules by \gem especially when substructures are not given. Furthermore, we observed that \gem fails to generate molecular graphs from scratch, when the model is trained on \zinc. We believe that this failure is due to the scarcity of valid graphs in large search space: for \zinc, shapes of input tensors are $38\times 10$ for feature tensors and $38\times 38\times 4$ for adjacency tensors.

One possible approach to overcome this limitation is to use lower dimensional continuous latent spaces as other generative approaches, such as \cite{simonovsky2018,jin2018,ma2018,decao2018}. Such latent spaces may enable more efficient search and higher validity, but, at the same time, hinder flexible substructure-preserving generation. Alternatively, \gem has room for introducing chemical rules to restrict its search space, such as inferring bond types of adjacency tensors from atom types of feature tensors. \gem uses the minimum prior knowledge for molecular graph generation, and thus, is potentially applicable for more general graph generation with subgraph preservation. We leave these possible improvements for future work.

\section{Conclusion}

In this paper, we have proposed \gem, an energy-based generative model for molecular graphs that can exactly preserve specified substructures, which is nearly impossible for existing approaches. \gem can design novel molecules with optimized properties, because energy function is explicitly instantiated. We have empirically demonstrate these abilities. Also importantly, \gem can be extended to general graph generation methods that can fix subgraphs.

By specifying available compounds as fixed substructures, \gem can design novel molecules that are expected to be easier to synthesize not only \emph{in silico} but also \emph{in vitro}. We hope \gem opens a new direction of de novo design.



\printbibliography
\end{document}

%% file: mysnipets.tex
\usepackage{amsmath,amsfonts,bm,amssymb,xspace,bbm,mathtools,color}

\def\1{\mathbbm{1}}

\def\vtheta{{\bm{\theta}}}

\def\vz{{\bm{z}}}

\def\vepsilon{{\bm \epsilon}}

\def\vtheta{{\bm \theta}}

\def\mJ{{\bm{J}}}

\DeclareMathAlphabet{\mathsfit}{\encodingdefault}{\sfdefault}{m}{sl}
\SetMathAlphabet{\mathsfit}{bold}{\encodingdefault}{\sfdefault}{bx}{n}
\newcommand{\tens}[1]{\bm{\mathsfit{#1}}}
\def\tA{{\tens{A}}}

\def\tH{{\tens{H}}}

\def\tU{{\tens{U}}}

\def\tX{{\tens{X}}}

\def\gB{\mathcal{B}}

\def\gD{\mathcal{D}}

\def\gG{\mathcal{G}}

\def\gM{\mathcal{M}}

\def\gZ{\mathcal{Z}}

\newcommand{\E}{\mathbb{E}}
\newcommand{\R}{\mathbb{R}}

\DeclarePairedDelimiterX{\infdivx}[2]{(}{)}{%
  #1\delimsize|\delimsize|#2%
}
\newcommand{\divergence}[3]{\ensuremath{D_{\mathrm{#1}}\infdivx{#2}{#3}}}
\newcommand{\KL}[2]{\divergence{KL}{#1}{#2}}

\DeclarePairedDelimiter{\abs}{\lvert}{\rvert}

\makeatletter
\DeclareRobustCommand\onedot{\futurelet\@let@token\@onedot}
\def\@onedot{\ifx\@let@token.\else.\null\fi\xspace}

\newcommand{\ie}{\emph{i.e}\onedot}
\newcommand{\eg}{\emph{e.g}\onedot}

\makeatother